\scrollmode

\documentclass[a4paper,12pt]{article}
\usepackage{epsfig}
\usepackage{amssymb}
\usepackage[a4paper]{geometry}

\begin{document}

\title{Towards a unification of {\sc hrt}\ and {\sc scoza}}

\author{A.~Reiner$^{1}$ and J.~S.~H\o ye$^2$,\\Teoretisk fysikk, Institutt
for fysikk,\\
Norges teknisk-naturvitenskapelige universitet (NTNU) Trondheim,\\
H\o gskoleringen 5, N-7491 Trondheim, Norway.\\
{}$^1$e-mail: {\tt areiner@tph.tuwien.ac.at}\quad
{}$^2$e-mail: {\tt Johan.Hoye@phys.ntnu.no}}

\maketitle

\begin{abstract}
The Hierarchical Reference Theory ({\sc hrt}) and the Self-Consistent
Ornstein-Zernike Approximation ({\sc scoza}) are two liquid state theories
that both furnish a largely satisfactory description of the critical
region as well as phase coexistence and the equation of state in
general.  Furthermore, there are a number of similarities that suggest the
possibility of a unification of
both theories.  As a first step towards this goal we consider the
problem of combining the lowest order $\gamma$ expansion result for
the incorporation of a Fourier component of the interaction with the
requirement of consistency between internal and free energies, leaving
aside the compressibility relation.  For simplicity we restrict
ourselves to a simplified lattice gas that is expected to display the
same qualitative behavior as more elaborate models.  It turns out that
the analytically tractable Mean Spherical Approximation is a solution
to this problem, as are several of its generalizations.  Analysis of
the characteristic equations shows the potential for a practical
scheme and yields necessary conditions any closure to the Ornstein
Zernike relation must fulfill for the consistency problem to be well
posed and to have a unique differentiable solution.  These criteria
are expected to remain valid for more general discrete and continuous
systems, even if consistency with the compressibility route is also
enforced where possible explicit solutions will require numerical
evaluations.\end{abstract}

\section{Introduction}

Both the Self-Consistent Ornstein-Zernike Approximation ({\sc scoza},
\cite{scoza:11,scoza:9,scoza:12}) and the Hierarchical Reference
Theory ({\sc hrt}, \cite{b:hrt:1}) have been found to give very accurate
results for fluids in thermal equilibrium.  In particular, the
respective non-linear parabolic partial differential equations ({\sc pde}s)
remain practical in the critical region, and their solution gives
non-classical, and partly Ising-like, critical indices.  The {\sc pde}s\
themselves are derived by obtaining the equation of state in two
independent ways and using thermodynamic consistency to fix a free
parameter in the direct correlation function.

Although both approaches thus appear similar in a number of aspects,
there are also marked differences: While both make use of the
compressibility route to thermodynamics, {\sc scoza}\ combines it with the
energy route expression for the \emph{internal} energy, whereas {\sc hrt},
inspired by momentum-space renormalization group theory, relies on
what might be called the fluctuation route to the \emph{free} energy
instead.  In addition to the density $\rho$, the independent variables
in the {\sc pde}s\ are therefore the inverse temperature $\beta = 1/k_B\,T$
and the momentum-space cutoff $Q$, respectively.  Starting from a
reference system of known properties at vanishing $\beta$ or at high
$Q$, the attractive interaction is then turned on by gradually
increasing its strength $\beta$ ({\sc scoza}), or by including its Fourier
component of wavenumber $Q$ at constant temperature until the
full system is recovered in the limit $Q \to 0$.

In the present contribution we want to investigate the possibility of
combining both approaches by imposing thermodynamic consistency of the
internal and free energies to obtain a differential formulation where
both the strength of the interaction and its Fourier components are
added successively.  This introduces an additional constraint so that
a further free parameter in the closure relation can be determined.
In combination with the compressibility route there are then two free
parameters that may be used, \textit{e.~g.{}}, to describe both the range
and the
amplitude of a contribution to the direct correlation function $c(r)$;
it might then be possible to describe the long-ranged tail that $c(r)$
is known to develop in the critical region.  However, our preliminary
investigations point to several difficulties with this approach near
phase coexistence.  The crucial problem is the vastly different
behavior of {\sc hrt}\ and {\sc scoza}\ in this part of the phase diagram:
{\sc hrt}\
has a solution at all densities, but spinodal and binodal coincide so
that the two phases coexist at infinite compressibility; {\sc scoza},
on the other hand, gives distinct binodal and spinodal curves but does not
have a
solution inside the spinodal.  At this point it is unclear how these
differences might be resolved by a simple modification of the direct
correlation function.  Even at temperatures above the critical one,
$\beta<\beta_c$, where full self-consistency should be attainable, we
will be confronted with differential equations that are of first order
in temperature and cutoff and of second order in the density.
Numerical solution of such a {\sc pde}\ can be demanding.

To simplify our task and gain some insight we here consider a
simplified lattice gas model, and we limit ourselves to combining the
energy and fluctuation routes without using the compressibility at
all.  Then there is only one free parameter left that can be
determined as a function of $\beta$ and $Q$.  After introducing our
model and establishing the {\sc pde}\ (section~\ref{sec:model}), we show
that consistency is achieved with the mean spherical approximation
({\sc msa}) and several generalizations of it (section~\ref{sec:msa}).  The
solution can therefore be given in closed analytic form, but the price
to pay for omission of the compressibility is the presence of the
{\sc msa}\ features, \textit{viz.{}}, ``irregular'' isotherms in the
critical region
and lack of a well-defined critical behavior for the lattice gas case
\cite{hoye:dr}.

In general, an analytical solution is also unlikely to be known
beforehand if other closures are used.  In that case one should seek
the solution numerically by integrating the equations for the
characteristic curves of the {\sc pde}.  Especially when the
compressibility relation not considered here is incorporated, a
numerical solution will be required, and we expect the properties of
the characteristics analyzed here to be useful more generally.  After
establishing relations and implications for more general situations
(section~\ref{sec:char:general}) we once more turn to the {\sc msa}\
equations.
It is found that the {\sc pde}\ admits a more general solution.  Usually,
the reference system also does not determine the solution at $Q=0$ for
finite temperature due to the behavior of the characteristics
(section~\ref{sec:char:msa}).  A generalized version of the {\sc pde}\ that
does
not make any assumptions on the form of the direct correlation
function of the system at cutoff $Q$ and inverse temperature $\beta$
other than that it depends on some unknown parameter function
$\lambda(\beta, Q)$ is considered in section~\ref{sec:char:oz}.  It is then
found that these deficiencies can be peculiar to the specific closure
employed.  Based on these findings, we finally state and consider two
complementary necessary conditions a parameterization of the solution
must fulfill for the consistency problem to be well posed and
physically reasonable (section~\ref{sec:conditions}).  These conditions
have the advantage that they can be checked in the high temperature
limit where significant simplifications are possible.  They are also
sufficiently general to apply to more realistic model systems, both
discrete and continuous.  In view of the relation between the
properties of a non-linear diffusion equation with two time-like
parameters on the one hand\ and the first-order {\sc pde}\ we consider in
the present
contribution on the other hand, the criteria found for the selection of
suitable
closure relations remain relevant even when the compressibility route
to thermodynamics is also taken into account.

\section{Adaptation of theory and model}

\label{sec:model}

\subsection{Model}

In the interest of simplicity, we here consider only a lattice gas (or
Ising model).  The main advantage of this choice is that the core
condition of vanishing pair distribution function $g(\vec r) = h(\vec
r)+1$ affects only the single point at $r\equiv\vert\vec r\vert=0$.  In
addition, we assume a
sufficiently long-ranged interaction so that the anisotropy imposed by
the lattice structure can be neglected.  Besides, an even more
long-ranged tail appears in the interaction for intermediate $Q$,
\textit{cf.{}}\
\cite{ar:8} and appendix~D.1 of \cite{ar:dr}.  When following the {\sc
hrt}\
recipe of successively including Fourier components of the
interaction, we will therefore do so without regard to the geometry of
the Brillouin zone.  In the same spirit we will often speak of finite
or infinite cutoffs and wavenumbers instead of specifying the
corresponding surfaces within the Brillouin zone explicitly
\cite{scoza:8,hrt:6}, and we will also write ${Q_{\infty}}$ for the maximum
cutoff in the calculation.  It is not expected that these
simplifications should affect the results qualitatively.  Furthermore,
the main conclusions of section~\ref{sec:conditions} are manifestly
independent of these assumptions.  

The system is assumed to differ by a perturbing attractive interaction
$-\psi(r)$, $\psi(r)>0$, from a reference system of known properties.
In the lattice case the simple hard core lattice gas can serve as the
reference system.  We can then use the inverse range $\gamma$ of the
attractive interaction as a perturbing parameter
\cite{hemmer:1964,lsb:1965}.  The zeroth order contribution past the
reference system is
the mean-field attractive van der Waals term, corresponding to a
pressure contribution of $-\frac12\,\tilde\psi(0)\,\rho^2$, where a tilde
indicates Fourier
transformed quantities.  In {\sc hrt}\ this corresponds to the zero-loop
diagram \cite{b:hrt:1}.  To next order in $\gamma$ Hemmer found
\cite{hemmer:1964}
\begin{equation} \label{hemmer:i}
I = -{\cal C}\,\int\ln\left(1-\tilde\mu(k)\tilde v(k)\right)\,{{\mathrm
d}}^3k \propto \gamma^3
\end{equation}
with
\begin{displaymath}\begin{array}{c}
\displaystyle {\cal C}     = \frac12\left(\frac1{2\pi}\right)^3,\\
\displaystyle \tilde v(k) = \beta\,\tilde\psi(k),\\
\displaystyle \beta  = \frac1{k_B\,T}.
\end{array}
\end{displaymath}
Here, $I$ is $-\beta$ times the first order
contribution to the Helmholtz free energy per unit volume, $k_B$ is
Boltzmann's constant, and $T$ is the temperature.  The integral is to
be extended over the Brillouin zone of the lattice that we here assume
to be the simple cubic one for which ${\cal C} \int{{\mathrm d}}^3k =
\frac12$.

The function $\mu(r)$ is related to the direct and total correlation
functions, $c_0(r)$ and $h_0(r)$, of the reference system
(indicated by the subscript~0) by
\begin{equation} \label{def:mu:ref}
\begin{array}{rl}
\displaystyle \tilde\mu(k) &\displaystyle= \rho+\rho^2\,\tilde h_0(k)\\
\displaystyle &\displaystyle=\frac\rho{1-\rho\,\tilde c_0(k)}.
\end{array}
\end{equation}
Here we have used the Ornstein-Zernike equation \cite{allg:26}
\begin{equation} \label{oz}
\begin{array}{c}
\displaystyle h(\vec r) = c(\vec r) + \rho\int h(\vec r-\vec r\,')\,c(\vec
r\,')\,{{\mathrm d}}^3r',\\
\displaystyle (1+\rho\tilde h(\vec k)) \, (1-\rho\tilde c(\vec k)) = 1.
\end{array}
\end{equation}
In the lattice case, the integral is to be replaced with a
sum.

In the limit $\gamma\to0$, the function $\tilde v(k)$ usually reduces to a
narrow peak of width $\gamma$ at $k=0$.  In this limit the
expression~(\ref{hemmer:i}) is the exact correction to the mean field
result.  In {\sc hrt}\ this corresponds to the infinite sum of one-loop
diagrams \cite{b:hrt:1}.

The interaction $\tilde v(k)$ may also have a narrow peak of width $\gamma$
at some other position, say, around $k=Q$.  In $r$ space this
corresponds to an infinitely weak but oscillating interaction with
wave vector $Q$, and eq.~(\ref{hemmer:i}) is still exact in the limit
$\gamma\to0$.  On the other hand, the renormalization procedure of {\sc
hrt}\ consists
of adding narrow pieces of $\tilde\psi$ of width $-{{\mathrm d}} Q>0$
around shorter
and shorter wave-vectors $Q$ at constant temperature.  With
${{\mathrm d}}^3k=4\pi\,k^2\,{{\mathrm d}} k$ and after division by
${{\mathrm d}} Q$, eq.~(\ref{hemmer:i})
then becomes
\begin{equation} \label{hrt:1}
\frac {\partial I}{\partial Q} = 4\pi\,{\cal
C}\,Q^2\,\ln\left(1-\tilde\mu(Q)\tilde v(Q)\right).
\end{equation}
This is also the first equation of the {\sc hrt}\ hierarchy, provided $\mu$
is derived from the structure of the system at slightly higher cutoff
$Q+\vert{{\mathrm d}} Q\vert$ rather than from the reference system at
${Q_{\infty}}$.
Eq.~(\ref{hrt:1}) is then formally exact, but it does not specify how
$\tilde\mu$ changes as Fourier components of $\tilde\psi$ are added and the
cutoff $Q$ approaches zero.  The formal answer to this problem is
provided by the higher equations of the {\sc hrt}\ hierarchy or some other
expansion for $\mu$.  In practical applications of {\sc hrt}\ only
eq.~(\ref{hrt:1}) is used, and the evolution of $\tilde\mu$ is determined
by
introducing a free parameter into the correlation functions and fixing
it by the requirement of consistency with the compressibility route.

One can also make a small change ${{\mathrm d}}\beta>0$ in the inverse
temperature and so effectively increase the interaction by an amount
$-{{\mathrm d}}\tilde v(k) = \tilde\psi(k)\,{{\mathrm d}}\beta$.  Taking
the limit ${{\mathrm d}}\beta\to0$ we
obtain
\begin{equation} \label{u1:dIdbeta}
-\rho\,u_1\equiv\frac {\partial I}{\partial\beta}
=
{\cal C}\,\int\frac{\tilde\mu(k)\,\tilde\psi(k)}{1-\tilde\mu(k)\,\tilde
v(k)}\,{{\mathrm d}}^3k.
\end{equation}
The quantity $u_1$ is nothing but the configurational internal energy
per particle beyond the zeroth order mean field term (besides a
self-energy term $-\frac12\,\psi(0)$ included here from the
$\rho$-term in $\mu$) due to a structure or correlation functions
given by
\begin{equation} \label{h:1stOrderGamma}
\rho+\rho^2\tilde h(k) = \frac{\tilde\mu(k)}{1-\tilde\mu(k)\,\tilde v(k)}.
\end{equation}
With $\mu$ given by eq.~(\ref{def:mu:ref}) this is exact only in the limit
$\beta\to0$ where the reference system is recovered.  At higher
$\beta$, it is only the first order correction in $\gamma$ for the
long-range part of the pair correlation function
\cite{hemmer:1964,lsb:1965}.  To higher order, or at finite $\beta$, it is
again not
obvious how $\tilde\mu(k)$ changes with temperature.  Here the {\sc scoza}\
recipe is to introduce some parameter into the closure that is also
fixed by the requirement of consistency with the compressibility
route.

\subsection{Specialization to lattice gas}

\label{sec:model:lg}

In the case of a lattice gas or the Ising model, the above expressions
are simpler.  The direct correlation function for the reference system
vanishes except at $r=0$ so that $\tilde\mu(k)$ is constant,
$\tilde\mu(k)\equiv\mu$; and from the pressure $p = -\ln(1-\rho)/\beta$ of
the hard core lattice gas serving as reference system we obtain its
value as
\begin{equation} \label{mu:const:hclg}
\mu = \tilde\mu(0)
\equiv
\left(\frac1\rho\,\frac{\partial\beta p}{\partial\rho}\right)^{-1}
=
\rho\,(1-\rho).
\end{equation}
As a simple approximation we can now replace the $\mu$ of
eq.~(\ref{mu:const:hclg}) by some effective value $\mu_e$ that may be
determined by imposing thermodynamic consistency between eqs.~(\ref{hrt:1})
and~(\ref{u1:dIdbeta}); in general $\mu_e$ will be a function of $\beta$
and
$Q$.  Such a change may cause the core condition $h(0)=-1$ to be
violated, which might be corrected by the introduction of an
additional parameter.  For the qualitative features we are interested
in here, however, the core condition is not expected to be important,
and we do not impose it below \cite{ar:4}.

As Fourier components are added to the interaction at successively
smaller cutoff $Q$, the interaction vanishes for $k<Q$ (except for the
mean field term at $k=0$), and the integral in eq.~(\ref{u1:dIdbeta}) must
be restricted to that part of the Brillouin zone where $k \ge Q$.
Furthermore, a small change ${{\mathrm d}} Q$ in the cutoff corresponds to
the
addition of a weak long-ranged oscillating tail in $r$ space, which
also contributes at $r=0$.  By setting the potential $\psi$ at $r=0$
equal to zero the corresponding unphysical contributions to the
internal energy from within the hard core can be avoided
approximately, and exactly in the limits $Q\to{Q_{\infty}}$ and $Q\to0$,
even when the core condition is not fulfilled.  With these
considerations and using spherical symmetry eqs.~(\ref{hrt:1})
and~(\ref{u1:dIdbeta}) become
\begin{equation} \label{dI:pde:msa}
\begin{array}{c}
\displaystyle \frac {\partial I}{\partial Q} = 4\pi\,{\cal
C}\,Q^2\,\ln\left(1-z\,\tilde\psi(Q)\right)\\
\displaystyle \frac {\partial I}{\partial\beta} = 4\pi\,\frac{{\cal
C}}\beta\,
\int_{k>Q}\frac{z\,\tilde\psi(k)}{1-z\,\tilde\psi(k)}\,k^2\,{{\mathrm d}} k
\end{array}
\end{equation}
with
\begin{equation} \label{def:mue}
z = \mu_e\,\beta.
\end{equation}
The above equations define a {\sc pde}\ for $\mu_e(\beta, Q)$ that is to be
solved for all $Q\ge0$, $\beta\ge0$.  In particular, the solution
should give the structural and thermodynamic properties of the target
system at $Q=0$ and all $\beta\ge0$.
In order to turn this into a well-defined problem we also have to
impose some boundary conditions.  The natural choice is to demand that
the reference system ($\mu_e=\mu$) should be recovered at $\beta=0$ and
at $Q={Q_{\infty}}$.  Whether this is sufficient to determine the solution
throughout the domain of the PDE, however, is not obvious and depends
not only on the general consistency problem but also on the particular
closure (the replacement of $\mu$ by $\mu_e$) and the toy model chosen.
We will return to this point in greater detail in section~\ref{sec:char}.

At first sight, an analytic solution of the given equations is not
obvious.  But as we will show in the following section, the {\sc msa}\
solution and generalizations thereof indeed fulfill eq.~(\ref{dI:pde:msa}).
On the other hand, the multiplicity of solutions found already indicates
that the
reference system boundary conditions (that are fulfilled in all cases)
do not determine a unique solution, \textit{cf.{}}\
sections~\ref{sec:msa:general}
and~\ref{sec:char:msa}.

\section{{\sc msa}\ and {\sc msa}-like solutions}

\label{sec:msa}

\subsection{Specific solution}

\label{sec:msa:pure}

The internal energy corresponding to the {\sc msa}\ correlation function
can be integrated explicitly to yield the free energy by utilizing
general {\sc msa}\ expressions \cite{scoza:12}.  The result is
\begin{equation} \label{I:msa}
I = -4\pi\,{\cal
C}\,\int_{k>Q}\ln\left(1-z\,\tilde\psi(k)\right)\,k^2\,{{\mathrm d}} k
+ J - \frac12\,\ln\left(1+2J\right).
\end{equation}
Here $J$ is the integral appearing in eq.~(\ref{dI:pde:msa}),
\begin{equation} \label{def:J}
\begin{array}{rl}
\displaystyle J&\displaystyle=\beta\,\frac {\partial I}{\partial\beta}\\
\displaystyle &\displaystyle=4\pi\,{\cal
C}\,\int_{k>Q}\frac{z\,\tilde\psi(k)}{1-z\,\tilde\psi(k)}\,k^2\,{{\mathrm
d}} k,
\end{array}
\end{equation}
and the core condition determines $z$ as
\begin{equation} \label{z:msa}
\begin{array}{rl}
\displaystyle z   &\displaystyle= \frac{\beta\mu}{1+2J}\\
\displaystyle \mu &\displaystyle= \rho\,(1-\rho).
\end{array}
\end{equation}
By differentiation of eq.~(\ref{I:msa}) with respect to $\beta$ and $Q$ and
using eqs.~(\ref{def:J}) and~(\ref{z:msa}) one finds that
eq.~(\ref{dI:pde:msa}) is
fulfilled.

The pair correlation function consistent with
solution~(\ref{I:msa})--(\ref{z:msa}) is obtained from
eq.~(\ref{h:1stOrderGamma}) by
replacing $\tilde\mu(k)$ by $\mu_e=z/\beta$ and restricting the interaction
to $k>Q$,
\begin{displaymath}
\rho + \rho^2\,\tilde h(k) = \frac{z / \beta}{1 -
z\,\tilde\psi(k)\,\Theta(k-Q)},
\end{displaymath}
where $\Theta$ is Heaviside's function.  Integrating the above over
the Brillouin zone and using eq.~(\ref{def:J}) one easily obtains
\begin{displaymath}
\rho + \rho^2\,h(0) = \frac z\beta\,(1+2J) = \mu = \rho-\rho^2,
\end{displaymath}
which shows that the core condition is fulfilled.

\subsection{More general solution}

\label{sec:msa:general}

The {\sc msa}\ solution just discussed is one specific solution of
eq.~(\ref{dI:pde:msa}).  A more general solution is suggested by previous
work on the resummation of $\gamma$ ordering for fluids
\cite{hoye:dr,hoye:nato}, a procedure that transforms the serious
divergence at the
mean field critical point into {\sc msa}\ type criticality while
maintaining self-consistency between the free and internal energies.

The generalized solution is then
\begin{equation} \label{general:I}
I = - {\cal C}\int_{k>Q}\ln\left(1-\mu_e\,\tilde v(k)\right)\,{{\mathrm
d}}^3k
- \sum_{n=1}^\infty \frac n{n+1}\,A_n\,K^{n+1}
\end{equation}
with ($z=\beta\mu_e$)
\begin{equation} \label{general:def:Khalf:mue}
\begin{array}{c}
\displaystyle K = \frac J\mu_e = {\cal C}\int_{k>Q}\frac{\tilde
v(k)}{1-\mu_e\,\tilde v(k)}\,{{\mathrm d}}^3k\\
\displaystyle \mu_e = \mu + \sum_{n=1}^\infty A_n K^n
\end{array}.
\end{equation}
Here the coefficients $A_n$ are independent of $\beta$ and $Q$ and
implicitly define $\mu_e$.  The $A_n$ must be fixed by some suitable
boundary conditions.  These must be imposed at finite temperature and
cutoff as $K$ vanishes at $\beta=0$ and at $Q={Q_{\infty}}$.  This shows
that
the reference system alone does not define a unique solution in the
present case, \textit{cf.{}}\ end of section~\ref{sec:model:lg}.  The {\sc
msa}\
solution~(\ref{I:msa}) corresponds to the more general
eq.~(\ref{general:I})
with one specific choice of the $A_n$.  By differentiation it is
easily verified that the above expression for $I$, like
expression~(\ref{I:msa}), solves both equations~(\ref{dI:pde:msa}).

\subsection{Wavenumber dependent direct correlation function}

\label{sec:msa:kdep}

Solution~(\ref{general:I}) can easily be extended to some very specific $k$
dependent $\mu_e$ and $z$.  Replacing $\mu_e$ by $\tilde\mu_e(k)$, the
Fourier
transform of $\mu_e(r)$ that now extends to outside the core, we can
write
\begin{equation} \label{kdep:mue}
\mu_e(r) = \mu\,\delta(\vec r) + \sum_{n=1}^\infty A_n(r)\,K(r)^n.
\end{equation}

This corresponds to graphs with $n$ parallel $K(r)$ bonds.  $K(r)$ is
the chain bond with renormalized hypervertex $\mu_e(r)$, and its
Fourier transform is
\begin{equation} \label{kdep:FK}
\tilde K(k) = \frac12\,\frac{\tilde v(k)}{1-\tilde\mu_e(k)\tilde
v(k)}\,\Theta(k-Q),
\end{equation}
the generalization of the integrand in eq.~(\ref{general:def:Khalf:mue}).
Eq.~(\ref{general:I}) now changes into
\begin{equation} \label{kdep:I}
I = - {\cal C}\int_{k>Q}\ln\left(1-\tilde\mu_e(k)\,\tilde
v(k)\right)\,{{\mathrm d}}^3k
- \sum_{n=1}^\infty \frac n{n+1}\,\int A_n(r)\,K(r)^{n+1}\,{{\mathrm
d}}^3r,
\end{equation}
and the {\sc pde}~(\ref{dI:pde:msa}) becomes
\begin{equation} \label{kdep:pde}
\begin{array}{rl}
\displaystyle \frac {\partial I}{\partial Q} &\displaystyle= 4\pi\,{\cal
C}\,Q^2\,\ln(1-\tilde\mu_e(Q)\,\tilde v(Q)),\\
\displaystyle \frac {\partial I}{\partial\beta} &\displaystyle=
4\pi\,\frac{{\cal C}}\beta\,
\int_{k>Q}\frac{\tilde\mu_e(k)\,\tilde v(k)}{1-\tilde\mu_e(k)\,\tilde
v(k)}\,k^2{{\mathrm d}} k.
\end{array}
\end{equation}
As mentioned at the beginning of section~\ref{sec:model}, we neglect the
lattice structure for our simplified model.  Of course, for a true
lattice system $\delta(\vec r)$ in eq.~(\ref{kdep:mue}) should be a
Kronecker $\delta$ at the origin, and the spatial integral in
eq.~(\ref{kdep:I}) should be a lattice sum.

The above solution is suggested by the work of H\o ye and Olaussen on
the two-dimensional Coulomb gas where the well-known
Kosterlitz-Thouless phase transition was evaluated on the basis of a
graph expansion \cite{ho:1980,ho:1981}.

Again it can be verified by differentiation that expression~(\ref{kdep:I})
solves eq.~(\ref{kdep:pde}) in the same way expression~(\ref{general:I})
solves
eq.~(\ref{dI:pde:msa}).  To do so one also needs the identity
\begin{displaymath}
\int f(r)\,g(r)\,{{\mathrm d}}^3r = 2{\cal C}\,\int \tilde f(k)\,\tilde
g(k)\,{{\mathrm d}}^3k.
\end{displaymath}

\section{Characteristics of the differential equations}

\label{sec:char}

\subsection{First-order {\sc pde}s\ and characteristics}

\label{sec:char:general}

So far we have only considered specific cases where exact solutions
are known.  In more general situations the characteristics can be
used.  These are curves along which the {\sc pde}\ of interest is
equivalent to a set of coupled ordinary differential equations.  In
the case of a first order {\sc pde}\ considered here, the solution can
always be obtained by integrating the characteristic equations,
starting from some point where the solution is known from some
boundary condition.  This is exact and does not incur any loss of
information or generality, and it generates a solution that is
necessarily differentiable along the characteristic.\footnote{See textbooks
on {\sc pde}s\ such as, \textit{e.~g.{}}, Leon Lapidus,
George F.~Pinder, \emph{Numerical solution of partial differential
equations in science and engineering}, New York (Wiley) 1982.}

Clearly, the solution along a full characteristic depends on only a
single point at the boundary.  This reflects the well-known capacity
of first-order {\sc pde}s\ for discontinuous solutions.  Another way in
which discontinuities may arise is from crossings of characteristic
curves: the solution then becomes multivalued, and a resolution
restoring uniqueness naturally leads to a \emph{shock} where the
solution is no longer differentiable.  A complementary difficulty one
may encounter are \emph{rarefactions}, \textit{i.~e.{}}, regions not
entered by any
characteristics.  In this case the solution is not defined there,
although there is often a natural weak formulation of the {\sc pde}\ that
can be used instead.

In the present work we consider consistency of the free and internal
energies at inverse temperature $\beta$ when gradually turning on
Fourier components of the interaction at ever smaller wavelength $Q$.
Neither shocks nor rarefactions are expected on physical grounds,
\textit{i.~e.{}},
the solution of the {\sc pde}\ should both exist and be differentiable for
$\beta\ge0$ and $Q\ge0$, at least away from the critical point and
outside the spinodal.  Using subscripts to denote partial derivatives,
a {\sc pde}\ like eq.~(\ref{dI:pde:msa}) is naturally written in the form
\begin{displaymath}
\begin{array}{rl}
\displaystyle \Psi_x(x,y) &\displaystyle= X(x,y;\lambda(x,y))\\
\displaystyle \Psi_y(x,y) &\displaystyle= Y(x,y;\lambda(x,y)).
\end{array}
\end{displaymath}
Here $\lambda$ is an unknown function of the independent variables $x$
and $y$.  As we will see, these equations imply a first order {\sc pde}\
for $\lambda$ the solution of which also gives the quantity of interest,
$\Psi$.

The above relations are not in the standard form
$F(x,y,\Psi,\Psi_x,\Psi_y)=0$ for a first order {\sc pde}.  The usual
expressions for the characteristics are thus not directly applicable.
One option is to invert, say, $X(x,y;\lambda)$ with respect to
$\lambda$.  The resulting expression for $\lambda$ can then be
inserted into the equation for $\Psi_y$, allowing the usual equations
for the characteristics of non-linear {\sc pde}s\ to be used.  The
disadvantage of this approach is that it leads to highly involved
expressions.

A simpler way of obtaining the characteristic equations is by cross
differentiation of the {\sc pde}: Setting $\Psi_{xy}=\Psi_{yx}$ gives
\begin{displaymath}
X_y + X_\lambda\,\lambda_y = Y_x + Y_\lambda\,\lambda_x.
\end{displaymath}
This is a quasi-linear first-order {\sc pde}\ for $\lambda$ of the form
$a\,\lambda_x + b\,\lambda_y = c$, for which the characteristic
equations are ${{\mathrm d}} x/a = {{\mathrm d}} y / b = {{\mathrm d}}
\lambda / c$.  In combination
with the {\sc pde}\ for $\Psi$ itself we immediately find the set of
characteristic equations,
\begin{equation} \label{general:char}
\frac{{{\mathrm d}} x}{Y_\lambda}
= -\frac{{{\mathrm d}} y}{X_\lambda}
=  \frac{{{\mathrm d}} \lambda}{X_y - Y_x}
=  \frac{{{\mathrm d}} \Psi_x}{X_x\,Y_\lambda - X_\lambda\,Y_x}
=  \frac{{{\mathrm d}} \Psi_y}{X_y\,Y_\lambda - X_\lambda\,Y_y}
=  \frac{{{\mathrm d}} \Psi}{\Psi_x\,Y_\lambda - \Psi_y\,X_\lambda}.
\end{equation}
At every point $(x, y)$ along a characteristic the ratios of these
differentials determine the direction of its tangent in the $(x, y)$
plane as well as the corresponding changes in $\lambda$, $\Psi_x$,
$\Psi_y$, and $\Psi$ along the curve.  The equations can easily be
solved numerically, \textit{e.~g.{}}, by predictor-corrector methods.  In
the
present contribution, however, we will not concern ourselves with
numerical evaluations.  Suffice it to say that a straightforward
implementation of these equations may require very small step sizes,
and that an accurate evaluation of the direction of the tangent may be
difficult, in the vicinity of points where $X_\lambda = Y_\lambda =
0$; discretizations treating $X_\lambda$ and $Y_\lambda$ as
perturbations relative to, \textit{e.~g.{}}, $X_{\lambda\lambda}$ and
$Y_{\lambda\lambda}$ can be employed there.

An important consequence of the equations given above concerns the
orientation of the characteristics in the $(x,y)$ plane at special
points: Evidently, wherever $X_\lambda=0$ and $Y_\lambda \ne 0$, the
tangent is parallel to the $x$ axis, \textit{i.~e.{}}, $y =
\mathrm{const}$.  By locating
the zeros of the $\lambda$ derivatives of the right hand sides of the
{\sc pde}\ we can thus rapidly gain a qualitative overview of the geometry
of the field of characteristic curves.  Considering the path of
integration in the $(\beta, Q)$ plane, pure {\sc hrt}\ obviously
corresponds to
characteristics at constant temperature, and pure {\sc scoza}\ to those at
constant cutoff.  We will therefore refer to these directions as
{\sc hrt}-like or {\sc scoza}-like, respectively.

For the {\sc pde}~(\ref{dI:pde:msa}), this directly implies the presence of
a
{\sc scoza}-like characteristic for $Q={Q_{\infty}}$ and of an {\sc
hrt}-like one at
$\beta=0$; together they cover that part of the domain where the
solution corresponds to the reference system.  In addition,
characteristics can have an {\sc hrt}-like tangent only in points where
$\tilde\psi(Q)=0$, and by considering all orders in the free parameter it
is seen that all characteristics have {\sc hrt}-like tangents whenever
$\tilde\psi(Q)$ vanishes at $Q<{Q_{\infty}}$ and $\beta>0$.

It is worth stressing that the significance of the characteristic
curves extends not only to the numerics: In addition to the
significance of shocks and rarefactions, neither of which are expected
in the present setting, the {\sc pde}\ is well-posed only when the
characteristics establish a one-to-one mapping from the points at
$Q=0$ to a subset of the boundary where the
solution and its first derivatives are known.  It is then possible to
start out from the boundary and to integrate along those curves until
the final solution is obtained at vanishing $Q$.

But even for a stable and well-posed {\sc pde}\ the characteristics are of
vital importance for the numerics, independently of whether the
discretization explicitly makes use of them: According to the Courant
Friedrichs Levy criterion \cite{numpde:cfl} any numerical scheme that
does not adhere to the flow of information represented by the
characteristics cannot be stable.

\subsection{Characteristics for the {\sc msa}-like case}

\label{sec:char:msa}

Even without prior knowledge of the generalized {\sc msa}\ solutions
presented in section~\ref{sec:msa}, some progress can be made towards the
solution of eq.~(\ref{dI:pde:msa}) by using the characteristics of the
{\sc pde}\ alone.  In the {\sc msa}\ case the unknown parameter function
$\lambda(x,y)$ of the preceding section corresponds to $\mu_e(\beta, Q)$.

\subsubsection{Explicit results}

\label{sec:char:msa:expl}

The situation is most transparent when considering only the limit of small
$z$ where eq.~(\ref{dI:pde:msa}) becomes to order
$O(z^2)$
\begin{displaymath}
\begin{array}{rl}
\displaystyle \frac {\partial I}{\partial Q}&\displaystyle=F'(Q)\,z,\\
\displaystyle \frac {\partial
I}{\partial\beta}&\displaystyle=\frac1\beta\,F(Q)\,z,\\
\displaystyle F(Q)&\displaystyle=4\pi\,{\cal
C}\int_{k>Q}\tilde\psi(k)\,k^2{{\mathrm d}} k.
\end{array}
\end{displaymath}
After elimination of $z$ and a change of variables from $Q$ to $F$ we
obtain a {\sc pde}\ for $I$ as a function of $F$ and $\beta$,
\textit{viz.{}},
\begin{equation} \label{I:pde}
F\,\frac {\partial I}{\partial F} - \beta\,\frac {\partial
I}{\partial\beta} = 0.
\end{equation}
The characteristics of this linear first-order {\sc pde}\ are determined
by
\begin{displaymath}
\frac{{{\mathrm d}} F}F
= -\frac{{{\mathrm d}}\beta}\beta
= \frac{{{\mathrm d}} I}0,
\end{displaymath}
which is trivially integrated to
\begin{equation} \label{I:pde:char}
\begin{array}{c}
\displaystyle F\,\beta = \mathrm{const},\\
\displaystyle I = \mathrm{const}.
\end{array}
\end{equation}
The general solution of the {\sc pde}\ is then any functional relation
between the two constants of integration above, \textit{i.~e.{}}, $I$ can
only
depend on $F(Q)\,\beta$,
\begin{equation} \label{I:pde:sol}
I = I(F(Q)\,\beta).
\end{equation}
This is also the case for the solutions of eq.~(\ref{dI:pde:msa}) given in
section~\ref{sec:msa} to first order in $z$.  Eqs.~(\ref{I:msa})
and~(\ref{general:I}) are both recovered to first order in $z$ by imposing
$I=0$ for $\beta=0$.

The situation is only slightly more complicated when terms of higher
order in $z$ are not neglected in eq.~(\ref{dI:pde:msa}).  We can obtain an
equivalent {\sc pde}\ by cross-differentiating the
equations~(\ref{dI:pde:msa})
and eliminating $\partial^2I/\partial Q\partial\beta$, as was done in
section~\ref{sec:char:general}.  From the relations~(\ref{dI:pde:msa}) and
the
definition~(\ref{def:J}) we find
\begin{displaymath}
-4\pi\,{\cal C}\,Q^2\,\frac{\tilde\psi(Q)}{1-z\,\tilde\psi(Q)}\,\left(\frac
{\partial z}{\partial\beta}\right)_Q
=\frac1\beta\left(\frac {\partial J}{\partial Q}\right)_\beta,
\end{displaymath} or \begin{displaymath}
\frac1z \left(\frac {\partial J}{\partial Q}\right)_z\,\left(\frac
{\partial z}{\partial\beta}\right)_Q
=\frac1\beta\left[\left(\frac {\partial J}{\partial Q}\right)_z
+ \left(\frac {\partial J}{\partial z}\right)_Q\,\left(\frac {\partial
z}{\partial Q}\right)_\beta
\right].
\end{displaymath}
Rearranging this yields a quasilinear first-order {\sc pde}\ for $z(\beta,
Q)$,
\textit{viz.{}},
\begin{equation} \label{z:pde}
\frac1z \left(\frac {\partial J}{\partial Q}\right)_z\,\left(\frac
{\partial z}{\partial\beta}\right)_Q
- \frac1\beta\,\left(\frac {\partial J}{\partial z}\right)_Q\,\left(\frac
{\partial z}{\partial Q}\right)_\beta
= \frac1\beta\,\left(\frac {\partial J}{\partial Q}\right)_z.
\end{equation}
As in eq.~(\ref{general:char}), the equation for the characteristics
becomes
\begin{displaymath}
\frac{{{\mathrm d}}\beta}{\displaystyle\frac1z\,\left(\frac {\partial
J}{\partial Q}\right)_z}
= -\frac{{{\mathrm d}} Q}{\displaystyle\frac1\beta\,\left(\frac {\partial
J}{\partial z}\right)_Q}
= \frac{{{\mathrm d}} z}{\displaystyle\frac1\beta\,\left(\frac {\partial
J}{\partial Q}\right)_z}.
\end{displaymath}
This immediately gives $z{{\mathrm d}}\beta = \beta{{\mathrm d}} z$ so that
\begin{equation} \label{z:pde:char:mue}
\mu_e=\frac z\beta = \mathrm{const}
\end{equation}
along the characteristics.  The relation between ${{\mathrm d}} Q$ and
${{\mathrm d}} z$
merely reproduces the total differential of $J(Q,z)$, implying
\begin{equation} \label{z:pde:char:J}
J=\mathrm{const}.
\end{equation}
The general solution of eq.~(\ref{z:pde}) is then any functional relation
between $z/\beta$ and $J$.  It is easily seen that this is consistent
with the expression for $\mu_e$ given in section~\ref{sec:msa:general};
constant $\mu_e=z/\beta$ and $J=\mu_e\,K$ also imply
constant $K$ so that $\mu_e$ must be a function of $K$ only, just as
indicated in eq.~(\ref{general:def:Khalf:mue}).

\subsubsection{Implications of the characteristics}

\label{sec:char:msa:interp}

As pointed out in section~\ref{sec:char:general}, the {\sc pde}\ has a
well-defined and unique solution only when the characteristics
establish a one-to-one mapping of the target system at various
temperatures ($Q=0$, $\beta>0$) onto points where the solution is
known from a suitable boundary condition.  For the {\sc
pde}~(\ref{dI:pde:msa}),
the most natural condition to impose is provided by the reference
system, the properties of which must be recovered both at $\beta=0$
and at $Q={Q_{\infty}}$.  On the other hand, among the general features of
this type of
{\sc pde}\ in section~\ref{sec:char:general} we found that these boundaries
also
coincide with the location of two characteristics.  Consequently there
can be no characteristic connecting a point on one of those boundaries
with a target system at non-zero $\beta$.  At first sight this seems
to render the reference system useless as a boundary condition.
Indeed, for the {\sc msa}-like solution considered here,
eq.~(\ref{z:pde:char:J}) shows that the reference system only determines
the solution where $J=0$, \textit{i.~e.{}}, on the boundary itself.

However, this is not necessarily a severe problem as $I(\beta, Q)$ is
expected to be continuous everywhere, including at the boundaries.
The reference system therefore still provides a valid description for
systems removed from the boundaries by a very small amount, where the
characteristic may have a direction different from that of the
boundary and may actually lead to the target system at non-zero
$\beta$.  In this case the reference system may still be used as a
boundary condition; an example where even an infinitesimal separation
is sufficient will be given in section~\ref{sec:char:oz}.  This situation
echos the one found in continuum fluids in {\sc hrt}: For those systems it
has always been necessary to start numerical integration of the
equations at some large but finite cutoff ${Q_{\infty}}$, typically on the
order of $10^2/\sigma$, where $\sigma$ is a length characteristic of
the repulsive hard core reference interaction \cite{ar:dr,ar:5}.
Neglecting the contribution of very high wavenumber fluctuations has
never been a practical problem, nor is this expected to be the case
here.

In order to investigate whether a purely numerical solution is
possible for the {\sc msa}-like equations considered in
section~\ref{sec:char:msa:expl}, let us consider the results obtained by
expanding the {\sc pde}\ to first order in $z$, valid in the limit of high
temperature.  For the moment assuming $\psi(0)\ne0$, there are
potentials such that $F(Q)$ vanishes only for $Q={Q_{\infty}}$.  In this
case
$I$ may be prescribed for all $\beta$ at some fixed $Q<{Q_{\infty}}$, the
characteristics stay at finite $\beta$ at all lower $Q$, and the
solution at $Q=0$ can be obtained at every temperature.

\begin{figure}[t]\vbox{\noindent\epsfbox{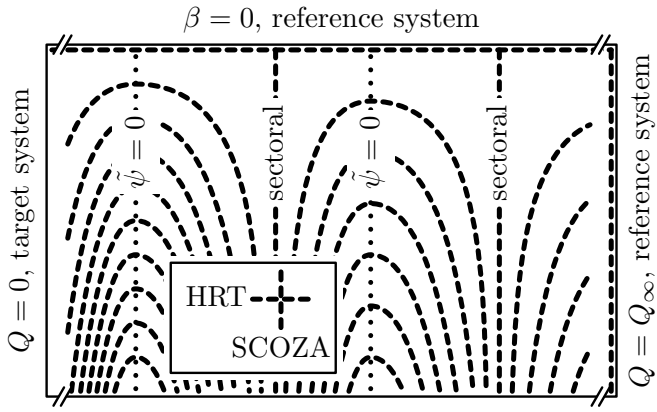}
\bigskip\caption{Sketch of the characteristics for the {\sc msa}-like
solution to first
order in $z$: Dashed lines show the characteristics whereas the
dotted lines give the locus of $\tilde\psi=0$ where the characteristics
have an {\sc hrt}-like tangent.  Sectorals are marked, as are the limits
where the target and reference systems are recovered and the
characteristics at $\beta=0$ and $Q={Q_{\infty}}$.  The inset gives the
directions corresponding to pure {\sc hrt}\ and {\sc scoza}, respectively.
There may or may not be a sectoral at $Q=0$, depending on whether
$\psi(0)$ vanishes or not.  As expected, there are no rarefactions,
nor are there crossings of characteristics except at
$\beta=0$.}\label{fig:sketch}}
\end{figure}

On the other hand, $F$ will be an oscillating function of $Q$ for many
short-ranged interactions, especially when the core condition is
approximately taken into account by demanding $\psi(0)=0$, \textit{cf.{}}\
section~\ref{sec:model:lg}.  At high temperature the field of
characteristics then qualitatively looks as sketched in
fig.~\ref{fig:sketch}.

An important but unfortunate feature of that sketch are the
{\sc scoza}-like characteristics starting at $\beta=0$ whenever $F(Q)=0$:
These \emph{sectorals}, as we shall call any characteristic the
tangent of which at $\beta=0+$ has a non-vanishing component in the
$\beta$ direction, divide the domain of the {\sc pde}\ into a
sequence of \emph{sectors}.  As there are no characteristics that
cross from one sector into a neighbouring one, no information is
transmitted across a sectoral, and the solutions in adjacent sectors
are largely independent.  The special significance of these characteristics
comes from the
fact that the solution for the target system ($Q=0$) is determined by
a boundary condition only when both lie in the same sector\footnote{This is
not the case for the analytical solution of
section~\ref{sec:msa:general} as constancy of the $A_n$ is
built right into eqs.~(\ref{general:I})
and~(\ref{general:def:Khalf:mue}).  It is then possible to
use a boundary condition to determine the $A_n$ in one
sector, and to use their values in another one.}.  Furthermore, the {\sc
scoza}-like orientation of a sectoral means that
the solution along it cannot depend on Fourier components of the
interaction at other $Q$ in the high temperature limit.

For lower temperature we have to use the full solution.  According to
eq.~(\ref{z:pde:char:J}), the sectorals are the curves where $J=0$.  As we
further follow along them, their direction will no longer be
{\sc scoza}-like, and the solution along them will depend on a range of
Fourier components.  Still, one of the properties that can be inferred
from the temperature dependence of the $J$ integral is that no
sectoral can ever get to $Q=0$ at non-zero $\beta$ since $J>0$ there.

This feature of the sectorals also conforms to our expectation for a
realistic closure relation.  For suppose that some sectoral starts at
$\beta=0$ and intermediate $Q > 0$ and finally reaches $Q=0$ at
$\beta=\beta_s > 0$.  Clearly it can only cover a finite $Q$ interval
so that the solution at $Q=0$, $\beta=\beta_s$ cannot depend on
Fourier components of the interaction outside this $Q$ range.  Absence
of crossings of characteristics immediately implies that the same is
also true for the properties of the target system at all $\beta <
\beta_s$.  This situation is clearly unacceptable except for
$\beta_s=0$.

Another consequence of the temperature dependence of the $J$ integral
is the absence of sectorals at given $Q$ for $\beta$ above some
$Q$-dependent limit; this limit is generally higher for larger $Q$.
There are then only a few classes of admissible configurations of the
characteristics in the $(\beta, Q)$ plane.  Most likely, the sectorals are
either driven towards infinite $\beta$ at ever higher $Q$, or pairs of
neighbouring sectorals join at their shared maximum $\beta$ and so
form loops; in the latter case, the maximum $\beta$ occurs at a cutoff
where $\tilde\psi(Q)=0$.  Unfortunately, both of these possibilities are
highly problematic: If sectorals stay separate but go towards infinite
$\beta$ and $Q$, there can be no finite temperature characteristic
connecting a point close to the reference boundary with one at $Q=0$,
$\beta>0$.  If, on the other hand, sectorals form loops, the same problem
will
arise at high cutoff.  In addition, loops demarcate regions that are
not entered by characteristics, implying rarefactions for $\beta\to0$.

\subsection{Connecting reference and target systems}

\label{sec:char:oz}

The properties of the characteristics just inferred for the {\sc msa}\ case
studied here are certainly unexpected and disappointing because they
do not allow us to go from the reference system at $Q={Q_{\infty}}$ to the
target system at vanishing $Q$.  We now show that this is not so much
a general defect of the combination of the energy and fluctuation
routes but a consequence of the specific parameterization of the
consistency problem.  To see this, let us consider an arbitrary
closure relation to the Ornstein-Zernike relation~(\ref{oz}), giving the
correlation functions $c$ and $g = h + 1$ for any combination of $Q$,
$\beta$, and some unknown parameter $\lambda(\beta, Q)$.  For example, the
{\sc msa}\ case considered so far corresponds to the direct correlation
function
\begin{displaymath}
\tilde c = \frac1\rho - \frac\lambda\mu + \Theta(k-Q)\,\tilde v,
\qquad \lambda = \frac\mu\mu_e.
\end{displaymath}
Another simple possibility is
\begin{displaymath}
\tilde c = \frac1\rho - \frac1\mu +
\left(\lambda+\Theta(k-Q)\right)\,\tilde v.
\end{displaymath}
This is suggested by the usual {\sc hrt}\ recipe for the continuum case
when the core condition is not explicitly taken into account
\cite{b:hrt:1,ar:4,hrt:4,hrt:10}.

In this more general situation with unspecified closure the correct
{\sc pde}\ is given by the earlier relations~(\ref{hrt:1})
and~(\ref{u1:dIdbeta}),
except that the energy integral must be restricted to $k>Q$ and that
$\tilde\mu$ now depends not only on $k$, $\beta$ and $Q$ but also on the
free parameter function $\lambda(\beta, Q)$.  As noted in
section~\ref{sec:model:lg}, the core condition must be
fulfilled at every $Q$ if the simple expression~(\ref{u1:dIdbeta}) is used,
or else there will be an unphysical contribution to the internal
energy from $r=0$.

In general, $\tilde\mu(k)$ is continuous at $k=Q$ whereas the direct
correlation function $\tilde c(k)$ of the system corresponding to cutoff
$Q$ has a discontinuity of height $\tilde v(Q)$ there.  The relation
between $\tilde\mu$ and $\tilde c$ is obtained from
eq.~(\ref{h:1stOrderGamma}) by
restricting the interaction to $k<Q$ again,
\begin{displaymath}
\rho+\rho^2\tilde h(k) = \frac{\tilde\mu(k)}{1-\tilde\mu(k)\,\tilde
v(k)\,\Theta(k-Q)}
\end{displaymath}
or
\begin{displaymath}
\tilde c(k) = \frac1\rho - \frac1{\tilde\mu(k)} + \tilde v(k)\,\Theta(k-Q).
\end{displaymath}
Without further specifying the closure, a number of properties of the
characteristics can easily be deduced: As the unknown parameter
$\lambda$ enters eq.~(\ref{dI:pde:msa}) only through $\tilde c$, we
immediately
see that the {\sc hrt}-like characteristic at $\beta=0$, the {\sc
scoza}-like
characteristic at $Q={Q_{\infty}}$, and the {\sc hrt}-like direction of the
tangents of all characteristics whenever $\tilde\psi(Q)=0$ remain in this
general setting.

Both for discrete and continuous systems the question is then whether
these boundary conditions at large but finite cutoff determine the
solution at $Q=0$.  We have just seen that this is usually not the
case for the {\sc pde}~(\ref{dI:pde:msa}).  As pointed out before, the
boundary
conditions must be imposed in the same sector where the target
system's properties are to be recovered.  For this to be possible and
the reference system to provide a valid initial condition for the
integration of the characteristic equations, it is necessary but not
sufficient that there be no sectorals above the initial value of $Q$,
nor between the boundary condition and the target system.

We then have to consider the existence and distribution of sectorals
once again, which crucially depends on the precise way in which the
free parameter enters the equations.  As an example, let
$\tilde c_\mathrm{ex}(k;\beta, Q)$ and $\tilde h_\mathrm{ex}(k;\beta, Q)$
be the direct and total
correlation functions for any solution of the {\sc pde}\ such as,
\textit{e.~g.{}}, any
of those presented in section~\ref{sec:msa}.  Inserting either of the two
sample closures
\begin{equation} \label{char:artifical:hrt}
\tilde c(k;\beta, Q,\lambda) = \tilde c_\mathrm{ex}(k;\beta, Q) +
\lambda\,\left(f_1(k)-f_1(Q)\right)
\end{equation}
and
\begin{equation} \label{char:artifical:scoza}
\tilde h(k;\beta, Q,\lambda) = \tilde h_\mathrm{ex}(k;\beta, Q) 
+
 \frac\lambda{\tilde\psi(k)}\,\left(f_2(2k-Q-{Q_{\infty}})-f_2(Q+{Q_{\infty}}-2k)\right)
\end{equation}
(with largely arbitrary functions $f_i$) into eqs.~(\ref{hrt:1})
and~(\ref{u1:dIdbeta}) yields a {\sc pde}\ for $\lambda(\beta, Q)$. 
Obviously, the
solution is $\lambda=0$ for all $\beta$ and $Q$ in either case so that
the original solution is reproduced, $\tilde c=\tilde c_\mathrm{ex}$ and
$\tilde h=\tilde h_\mathrm{ex}$.  According to eq.~(\ref{general:char}),
however, for
eq.~(\ref{char:artifical:hrt}) the characteristics are {\sc hrt}-like
everywhere
(${{\mathrm d}}\beta=0$, so that there are no sectorals at all), whereas
they are
all {\sc scoza}-like everywhere for eq.~(\ref{char:artifical:scoza})
(${{\mathrm d}} Q=0$,
so that there are infinitely many sectorals).  Artificial and
impractical as these examples are, they demonstrate that two different
parameterizations of one and the same solution may have vastly
different consequences, and eq.~(\ref{char:artifical:hrt}) in particular
shows the potential for a closure free of sectorals.

\section{Necessary conditions for well-posedness}

\label{sec:conditions}

We therefore conclude that the difficulties encountered in the case
studied in 
section~\ref{sec:char:msa} are not inherent in the consistency problem but
merely a consequence of the particular {\sc msa}-like closure used.  The
results of our investigation thus leave open the possibility of a
practical scheme, drawing our attention to the precise
parameterization of the solution.

While formulation of practical closure relations to be used together
with eq.~(\ref{dI:pde:msa}) lies outside the scope of the present
contribution, we can give two general non-trivial conditions an ansatz
for $\tilde c$ must fulfill for the consistency problem to be well-posed
and for a unique and differentiable solution to exist at all
$\beta\ge0$, $Q\ge0$: (1)~There must not be any sectorals in the sense
of the definition given in section~\ref{sec:char:oz}; and (2)~any
characteristic passing through a state with $0<\beta=O(\epsilon)$ at
some $Q>0$ must remain at a strictly positive inverse temperature
$\beta$ of order $O(\epsilon)$ at lower cutoffs.  While the former is
quite obvious in the light of section~\ref{sec:char:msa:interp}, the second
condition has not featured prominently so far: It merely expresses the
absence of rarefactions and shocks at infinite temperature, and for
the {\sc msa}\ case as treated here these already imply sectorals,
\textit{cf.{}}\
section~\ref{sec:char:msa:interp}.

Both of these criteria share the advantage of involving only the limit
$\beta\to0$.  They can therefore be checked by low order expansions in
$\beta$, which vastly simplifies analysis of the suitability of some
specific closure relation in the context of eq.~(\ref{dI:pde:msa}).

Another important property is that both of them are quite general: Not
only do they apply to discrete and continuous systems equally, they
are also independent of the particular simplifications we chose to
make in the present contribution.  In particular, a closure may give a
solution where the core condition is violated, leading to a spurious
contribution to the energy integral of eq.~(\ref{u1:dIdbeta}).  In this
case more care should be exercised when evaluating the internal
energy, leading to a modification of the {\sc pde}\ \cite{ar:13}.  The two
conditions set out above, however, remain equally valid nevertheless.

A careful study of the characteristics, especially of their compliance
with the criteria just set out, is thus of vital importance for the
consistency problem involving the energy and fluctuation routes only.
The results of such an analysis are also highly relevant when the
compressibility route enters the picture.  For suppose that
$I(\beta, Q,\rho)$ is the solution of the more elaborate {\sc pde}\ so
obtained.
By evaluating $\partial^2I/\partial\rho^2$ and inserting this into the
compressibility sum rule used in both pure {\sc hrt}\ and pure {\sc scoza}
we
arrive at a constraint involving both of the two unknown parameters in
the parameterization of the solution at any $(\beta, Q,\rho)$.  Restriction
to fixed density in the spirit of the line method then again leads to
a problem of the type considered here, and the properties of the
characteristics of this restricted problem must be taken into account
when solving, or discretizing, the {\sc pde}.  In particular, the density
dependence of the directions of the characteristics in the $(\beta, Q)$
plane are then of prime importance for the numerical tractability of
the {\sc pde}\ by finite difference methods.  At this point, however, it is
not clear how to handle the situation where $\beta$ does not increase
monotonously along the characteristics, nor do we know whether that
case actually occurs with physically plausible closure relations and
interactions.

In summary, in the present contribution we have introduced and started
to tackle the problem of unifying {\sc scoza}\ and {\sc hrt}, pointing out
the
importance of the condition of consistency between free and internal
energies and for the time being setting aside consistency with the
isothermal compressibility.  A simplified lattice model system allowed
us to derive a suitable {\sc pde}\ and to study its known analytical
solutions of {\sc msa}\ type.  In general, however, we cannot assume an
analytical solution to be available, and we therefore then turned to
the problem of solving the {\sc pde}\ without previous knowledge of the
exact solution, especially with a view towards a numerical
implementation: This highlighted the importance of the characteristic
curves of the {\sc pde}, and we have studied some of their properties and
consequences.

\section*{Acknowledgments}

{\sc ar}\ gratefully acknowledges financial support from \textit{Fonds zur
F\"orderung der wissenschaftlichen Forschung (FWF)} under
projects~P15758-N08 and~J2380-N08.

\end{document}